\documentclass[a4paper]{article}
\usepackage{CJK}
\usepackage{amsmath,amsthm}
\usepackage{amsmath}
\usepackage{amssymb}
\usepackage{amsfonts}
\usepackage{eufrak}
\usepackage{eucal}
\usepackage{fancyhdr}
\usepackage{graphicx}
\usepackage{float}
{\setlength\arraycolsep{2pt}
\usepackage{mathrsfs}
\usepackage{graphicx,fancyhdr}
\usepackage{graphicx} 
\usepackage{CJKnumb}
\usepackage{titlesec}
\usepackage{amsthm}
\usepackage{cite}
\usepackage[utf8]{inputenc}
\usepackage[colorlinks,urlcolor=blue,citecolor=blue,linkcolor=blue]{hyperref}
\usepackage{chngcntr}
\usepackage{makecell}
\usepackage{tabularx}

\counterwithout{figure}{section}%取消图片按章节编号

\usepackage{flushend} % 参考文献双栏平均
\usepackage{ulem}    % 参考文献去掉下划线
\usepackage[numbers, sort]{natbib} % 保证参考文献不会出现乱序

\begin{document}
\setcounter{page}{1}
\title{Quark Nova with the Producing of Color-Flavor Locked Quark Matter}
\author{Jia-rui Guo \\ NanKai University, Tianjin 300071, China
\thanks{Email address: 2010213@mail.nankai.edu.cn}
}
\maketitle
%\address{NanKai University, Tianjin 300190, China\\ }

\begin{abstract}
In this paper, we suggest that the process in quark nova explosion may exist widely in various kinds of supernova, although it only happens in a small part in the core in most cases. And the contribution to the energy releasing of whole supernova explosion can also be provided by QCD interacting term. We use Cornell Potential to describe it with its main part provided by confirm in the condition we focus of. In this way we derive a general equation of energy quantity to be released in quark nova process related to several parameters, and we also give the result with gravitational effect being considered. After quark nova explosion process, the remnant can be a quark star, or a neutron star with quark matter core if this process only happens in a small part inside the compact star instead of a “full” quark nova. We will also use a more generalized approach to analyse the strangelets released from quark nova and will draw a possible interpretation of why effects caused by strangelets have not been observed yet. Our result suggests that the ordinary matter can only spontaneously transform into strange quark matter by crushing them into high pressure under the extreme condition in compact star, although generally the reaction would really be exergonic.

\end{abstract}

\section{Introduction} \label{sec1}
Quark nova is a secondary collapsing after neutron stars preliminary formed and it can possibly  give a considerable effect to the whole supernova event. It releases energy because the quark matter state is believed to be the ground state of matter  \cite{witten1984cosmic}. In an environment of high density and high pressure, the quark matter is believed to be in Color-Flavor Locked (CFL) phase in the compact stars  \cite{alford1999color}. In the previous studies of supernova explosion models with ordinary simulations, the energy obtained from the explosion itself is not enough for pushing the outer layer of the star to escape from its gravity. R. Ouyed, J. Dey, and M. Dey proposed the quark nova hypothesis, which suggested that the phase transformation to quark matter will release much energy  and gave estimation about the order of magnitude of energy released in this process \cite{ouyed2002quark}. In this paper, we infer the process of quark nova explosion can also happen in supernovas with producing of neutron star, although there may be only a small part of neutron star matter transforms into quark matter in this case. This could suggest a quark-nova solution for the problem of the ordinary supernova explosion model.  One possible way to verify if the quark nova hypothesis is the solution to this problem is by checking if the energy released from quark nova explosion can satisfy the requirement for outer layer of the star to escape from gravity of the star and release enough radiation which can match with known data. In that case, we compare the differences of the conditions between quark matter (especially in Color-Flavor Locked (CFL) phase  \cite{alford1999color}) and nuclei matter, applying this difference in quark nova calculating, and comparing the result with the known data. A problem of this hypothesis is where the strangelets produced by quark nova have gone. To explain that, result of part \ref{sec5} suggests strangelets could only naturally form under great pressure deeply inside compact stars, so most of strangelets might not be able to escape from the compact star, and the strangelets may not be able to cause the ordinary matter spontaneously transform into strange quark matter by crushing them into high pressure in the common condition outside the compact star as well. Natural Unit System will be used everywhere in this paper unless there is a special declaration. 

\section{Background Analysis for Strange Quark Matter} \label{sec2}

In this part, we will give a general analysis about the background condition (such as quantum chromodynamics, thermodynamics and space-time metric) of quark matter, to prepare for the following parts. In the condition of quark nova, the pressure and chemical potential will increase greatly and crush ordinary nuclei matter into quark matter. In this process, different QCD phases will be produced, like 2SC \cite{blaschke2000diquark}, non-CFL strange quark matter (u, d, s, e) and CFL, therefore we will discuss these phases and their equations of state. Here we will analysis it ignoring whether it is a quark star (or quark matter core) or just a strangelet (which can be produced by quark nova), because they would all satisfy the same rule. Given the complete Hamiltonian of 2SC phase

\begin{equation} \label{e1}
\begin{aligned}
\mathcal{H}_{2  {SC}}=&-\frac{1}{4}\left(\partial_{\mu}  {A}_{v}^{\alpha}-\partial_{v} A_{\mu}^{\alpha}\right)\left(\partial^{\mu}  {A}^{\alpha v}-\partial^{v}  {~A}^{\alpha \mu}\right)-\frac{\left(\partial^{\mu}  {A}_{\mu}^{\alpha}\right)^{2}}{2 \alpha} \\
+&\bar{\psi}_{ {i}}\left( {i} \gamma^{\mu} \partial_{\mu}- {m}\right) \psi_{ {i}}-\left(\partial^{\mu}  {C}^{\alpha^{*}}\right)\left(\partial_{\mu}  {C}^{\alpha}\right)
+\mathcal{H}_{\text {electron }}+\mathcal{H}_{\text {(QED)int }}\\
+&  {n}_{ {q}}\left(-\frac{6 \Delta^{2} \mu^{2}}{n_{q} \pi^{2}}+\varepsilon_{( {QCD})  {int}}\right)
\end{aligned}
\end{equation}

Here $\varepsilon_{(QCD)int}$ means the QCD interacting potential energy per quark, and in this part of this paper, $n_q$ stands for the number density of quarks. The QCD interaction in 2SC phase mainly comes from the scattering of $u, d$ quarks and gluons, with the scattering amplitude given by

\begin{equation}
 {i} \mathcal{M}= {T}_{ {ji}}^{\alpha}  {T}_{ {kl}}^{\alpha}\left( {ig}_{ {s}}\right)^{2} \overline{ {u}}_{ {j}}\left( {p}_{2}\right) \gamma^{\mu}  {u}_{ {i}}\left( {p}_{1}\right) \frac{ {i}\left[- {g}_{\mu v}+(1-\xi) \frac{ {k}_{ {\mu}}  {k}_{ {v}}}{ {k}^{2}}\right]}{ {k}^{2}}  {u}_{ {k}}\left( {p}_{3}\right) \gamma^{v} \overline{ {u}}_{1}\left( {p}_{4}\right) \label{e2}
\end{equation}
According to above, we can derive the equation of state for this condition by the following

\begin{equation} \label{e3}
\begin{gathered}
 {PV}=-\int  {d}^{3}  {x} \mathcal{H}_{2  {SC}}+\bar{\Psi}_{ {i}}\left( {i} \gamma^{\mu} \partial_{\mu}- {m}\right) \psi_{ {i}}  {n}_{ {q}}+ {T} \int_{0}^{ {T}}\left( {C}_{ {V}}-\frac{3  {P}^{2}  {r}^{6}}{2  {a}_{ {s}}  {n}_{ {q}} \varepsilon_{( {QCD})  {int}}  {T}}\right)  {dT} \\
+\frac{1}{4}  {~g}^{\mu v}  {~F}^{2} 
- {F}^{\mu \alpha}  {F}_{\alpha}^{\mu }   
\end{gathered}
\end{equation}

When transformed into strange quark matter, the complete Hamiltonian is 
\begin{equation} \label{e4}
%\begin{aligned}
\begin{gathered}
\mathcal{H}_{ {sqm}}=-\frac{1}{4}\left(\partial_{\mu}  {A}_{ {v}}^{\alpha}-\partial_{ {v}}  {A}_{\mu}^{\alpha}\right)\left(\partial^{\mu}  {A}^{\alpha v}-\partial^{ {v}}  {A}^{\alpha \mu}\right)-\frac{\left(\partial^{\mu}  {A}_{\mu}^{\alpha}\right)^{2}}{2 \alpha}+\bar{\psi}_{ {i}}\left( {i} \gamma^{\mu} \partial_{\mu}- {m}\right) \psi_{ {i}}-\left(\partial^{\mu}  {C}^{\alpha^{*}}\right)\left(\partial_{\mu}  {C}^{\alpha}\right) \\
+\mathcal{H}_{\text {electron }}+\mathcal{H}_{\text {(QED)int }}+ {E}_{\text {sym }} \\
+ {n}_{ {q}}\left(-\frac{3 \Delta_{\mathrm{ds}}^{2} \mu^{2} \eta}{\mathrm{n}_{\mathrm{q}} \pi^{2}}-\frac{3 \Delta_{\mathrm{us}}^{2} \mu^{2} \eta}{\mathrm{n}_{\mathrm{q}} \pi^{2}}-\frac{3(1-2 \eta) \Delta_{\mathrm{ud}}^{2} \mu^{2}}{\mathrm{n}_{\mathrm{q}} \pi^{2}}+\varepsilon_{( {QCD})  {int}}\right)
\end{gathered}
%\end{aligned}
\end{equation}

Here $\eta$ is the porportion of s quark and satisfies $0<\eta \leq \frac{1}{3}$
Specially, when $\eta =\frac{1}{3}$, it would transform to CFL phase. 

\begin{equation} \label{e5}
\begin{aligned}
\mathcal{H}_{ {CFL}}=&-\frac{1}{4}\left(\partial_{\mu}  {A}_{ {v}}^{\alpha}-\partial_{v}  {~A}_{\mu}^{\alpha}\right)\left(\partial^{\mu}  {A}^{\alpha \nu}-\partial^{v}  {~A}^{\alpha \mu}\right)-\frac{\left(\partial^{\mu}  {A}_{\mu}^{\alpha}\right)^{2}}{2 \alpha}\\
+&\bar{\psi}_{ {i}}\left( {i} \gamma^{\mu} \partial_{\mu}- {m}\right) \psi_{ {i}}-\left(\partial^{\mu}  {C}^{\alpha^{*}}\right)\left(\partial_{\mu}  {C}^{\alpha}\right) \\
+&\mathcal{H}_{( {QED})  {int}}+ {n}_{ {q}}\left(-\frac{6 \Delta_{\mathrm{CFL}}^{2} \mu^{2}}{\mathrm{n}_{\mathrm{q}} \pi^{2}}+\varepsilon_{( {QCD})  {int}}\right)
\end{aligned}
\end{equation}

In order to calculate the symmetry energy, we expand the binding energy per baryon number in isospin asymmetry $\delta=\frac{3 \eta}{1-\eta}$, then we have

\begin{equation} \label{e6}
  {E}\left(  {n}_{  {B}}, \delta,   {n}_{  {s}}\right)=  {E}\left(  {n}_{  {B}}, \delta=0,   {n}_{  {s}}\right)+\left.\sum_{  {n}=1,2,3 \ldots} \frac{\delta^{  {n}}}{  {n} !} \frac{\partial^{  {n}}   {E}\left(  {n}_{  {B}}, \delta,   {n}_{  {s}}\right)}{\partial \delta^{  {n}}}\right|_{\delta=0}
\end{equation}

If the Coulomb interaction among quarks is weak enough to be ignored, the exchange symmetry between u and d quarks will then remove the terms which have an odd number as the value of $n$. Because the higher-order terms are very small, we can derive symmetry energy by the order of $n=2$, which means

\begin{equation} \label{e7}
  {E}_{  {sym}}\left(  {n}_{  {B}}, \delta,   {n}_{  {S}}\right)=\left.\frac{1}{2} \frac{\partial^{2}   {E}\left(  {n}_{  {B}}, \delta,   {n}_{  {s}}\right)}{\partial \delta^{2}}\right|_{\delta}
\end{equation}

Here $  \mathrm{E}\left(\mathrm{n}_{\mathrm{B}}, \delta, \mathrm{n}_{\mathrm{s}}\right)=3 \mathrm{P}+4 \mathrm{~B}+\frac{-\left(6 \eta \Delta_{\mathrm{ds}}^{2}+6 \eta \Delta_{\mathrm{us}}^{2}+6(1-2 \eta) \Delta_{\mathrm{ud}}^{2}\right)+3 \mathrm{~m}_{\mathrm{S}}^{2}}{2 \pi^{2}} \mu^{2}$ with $\eta=\frac{\delta}{3+\delta}$. 

And the equation of state for this condition

\begin{equation}
\begin{gathered}
 {PV}=-\int  {d}^{3}  {x} \mathcal{H}_{CFL}+\bar{\Psi}_{ {i}}\left( {i} \gamma^{\mu} \partial_{\mu}- {m}\right) \Psi_{ {i}}  {n}_{ {q}}+ {T} \int_{0}^{ {T}}\left( {C}_{ {V}}-\frac{3  {P}^{2}  {r}^{6}}{2  {a}_{ {s}}  {n}_{ {q}} \varepsilon_{( {QCD})  {int}}  {T}}\right)  {d}  {T}\\+ \frac{1}{4}  {~g}^{\mu v}  {~F}^{2} 
- {F}^{\mu \alpha}  {~F}_{\alpha}^{\mu }   \label{e8}
\end{gathered}
\end{equation}

C above is heat capacity given by $C_{\mathrm{V}}=\frac{1}{\mathrm{u}^{2}} \frac{\partial \Omega}{\partial \mathrm{T}^{2}}$,u represents the speed of sound, which follows
\begin{equation}
\mathrm{u}^{2}=\frac{\mathrm{p}_{\mathrm{F}}^{2}}{3 \mathrm{~m}_{\mathrm{q}}^{2}}\left[1+\frac{2 \mathrm{ap}_{\mathrm{F}}}{\pi}+\frac{8(11-2 \ln 2)}{15 \pi^{2}}\left(\mathrm{ap}_{\mathrm{F}}\right)^{2}\right] \label{e9}
\end{equation}
$\mathrm{m}_{\mathrm{q}}$ is quark mass, a is the scattering length,$\mathrm{p}_{\mathrm{F}}$ is Fermi momentum.

Equation \eqref{e3} and \eqref{e8} is about idealized quark matter without considering other effects in macroscopic compact star structure like pressure between different regions or gravity. Here we use a simplification to estimate $\varepsilon_{(Q \mathrm{CD}) \text { int }} .$ Generally, the color confinement limits the distance between interacting quarks and the high quark number density in CFL quark matter which makes it reasonable to consider the existence of a uniform color and flavor neutrality in this region. Thus, the potential energy of interaction between a random quark and its environment can be simplified as interacting with its anti-quark. This is because if we consider an effective "quark" to interact with the quark instead of the environment around but except this quark, this effective "quark" requires an opposite position, charge, color and so on as a balance to keep the neutrality. The Cornell Potential \cite{eichten1975spectrum} between the quark and the effective "(anti)quark" is therefore written as

\begin{equation} \label{e10}
 {V}=-\frac{4  {a}_{ {s}}}{3  {r}}+ {br}+ {c}
\end{equation}

When the CFL quark matter is crushed into higher quark number density, consider virial theorem we have
\begin{equation} \label{e11}
\left(\frac{\partial  {H}_{ {kin}}}{\partial  {r}}\right)_{ {N}_{ {q}}}=-\frac{1}{2}\left(\frac{\partial  {H}_{ {int}}}{\partial  {r}}\right)_{ {N}_{ {q}}}
\end{equation}
$ {N}_{ {q}}$ is total quark number, and $ {r}$ is average distance between quarks. This implies the temperature of CFL quark matter will naturally increase when it is crushed denser, and the pressure will increase with it, stop the quark matter from an unlimited collapsing. To add up the gravitational effect, we write the energy-momentum tensor of CFL quark matter with QED contribution considered as

\begin{equation} \label{e12}
\begin{aligned}
& {T}_{ {CFL}}^{\mu  {v}}=\left[\bar{\psi}_{ {i}}\left( {i} \gamma^{\mu} \partial_{\mu}- {m}\right) \psi_{ {i}}\right]  {U}^{\mu}  {U}^{ {v}}+ {p}\left( {g}^{\mu \nu}+ {U}^{\mu}  {U}^{v}\right)-\left(\partial^{\mu}  {C}^{\alpha^{*}}\right)\left(\partial_{\mu}  {C}^{\alpha}\right) \\
&-\frac{1}{4}\left(\partial_{\mu}  {A}_{ {v}}^{\alpha}-\partial_{v}  {~A}_{\mu}^{\alpha}\right)\left(\partial^{\mu}  {A}^{\alpha v}-\partial^{ {v}}  {A}^{\alpha \mu}\right)-\frac{\left(\partial^{\mu}  {A}_{\mu}^{\alpha}\right)^{2}}{2 \alpha} \\
&+ {n}_{ {q}}\left(-\frac{6 \Delta_{\mathrm{CFL}}^{2} \mu^{2}}{\mathrm{n}_{\mathrm{q}} \pi^{2}}+\varepsilon_{( {QCD})  {int}}\right)+\frac{1}{4 \pi}\left( {F}_{ {QED}}^{\mu \sigma}  {F}_{\sigma( {QED})}^{ {v}}\right. \\
&\left.-\frac{1}{4}  {~g}^{\mu v}  {~F}_{\alpha \beta( {QED})}  {F}_{ {QED}}^{\alpha \beta}\right)
\end{aligned}
\end{equation}

The pressure here is given in the equation of state above, and $ {U}^{\mu}$ is the unit spacelike vector in the radial direction. Solve Einstein's field equation of CFL quark matter, consider the quark star or quark matter core of neutron star to be spherically symmetric, we can derive the space-time metric

\begin{equation} \label{e13}
\begin{gathered}
\mathrm{d} \mathrm{s}^{2}=-\left[1-\frac{2 \int \mathrm{d}^{3} \mathrm{x} \frac{\mathrm{n}_{\mathrm{q}}\left(\mathrm{m}_{\mathrm{u}}+\mathrm{m}_{\mathrm{d}}+\mathrm{m}_{\mathrm{s}}\right)}{3}}{\mathrm{r}}+\frac{\mathrm{Q}^{2}}{\mathrm{r}^{2}}-2 \int \mathrm{d}^{3} \mathrm{x} \frac{\mathrm{n}_{\mathrm{q}}\left(-\frac{4 \mathrm{a}_{\mathrm{s}}}{3 \mathrm{r}}+\mathrm{br}+\mathrm{c}-\frac{6 \Delta_{\mathrm{CFL}}^{2} \mu^{2}}{\mathrm{n}_{\mathrm{q}} \pi^{2}}\right)}{\mathrm{r}}\right] \mathrm{dt}^{2} \\
+\left[1-\frac{2 \int \mathrm{d}^{3} \mathrm{x} \frac{\mathrm{n}_{\mathrm{q}}\left(\mathrm{m}_{\mathrm{u}}+\mathrm{m}_{\mathrm{d}}+\mathrm{m}_{\mathrm{s}}\right)}{3}}{\mathrm{r}}+\frac{\mathrm{Q}^{2}}{\mathrm{r}^{2}}\right. \\
\left.-2 \int \mathrm{d}^{3} \mathrm{x} \frac{\mathrm{n}_{\mathrm{q}}\left(-\frac{4 \mathrm{a}_{\mathrm{s}}}{3 \mathrm{r}}+\mathrm{br}+\mathrm{c}-\frac{6 \Delta_{\mathrm{CFL}}^{2} \mu^{2}}{\mathrm{n}_{\mathrm{q}} \pi^{2}}\right)}{\mathrm{r}}\right]^{-1} \mathrm{dr}^{2}+\mathrm{r}^{2} \mathrm{~d} \Omega^{2}
\end{gathered}
\end{equation}

where $Q$ means the total charge, and $\overline{ {r}}$ is the average effective interacting distance between the quarks in this quark matter. Calculating with the previous research \cite{bali2001qcd}, it would cause a considerable effect on the gravitational potential of quark star. Figure 1 show the exact result of the correction by this effect, and the following parts will use an equivalent mass $ {G}$. 

To derive $G$ we write the time-like killing field $\xi^{\alpha}$ in space-time with a metric of equation (13), therefore the gravitational acceleration is given by

\begin{equation} \label{e14}
\begin{aligned}
&|\vec{g}|=\frac{2 \int d^{3} x\left[\frac{n_{q}\left(m_{u}+m_{d}+m_{s}\right)}{3}-\frac{4 a_{s}}{3 r}+b r+c-\frac{6 \Delta_{C F L}^{2} \mu^{2}}{n_{q} \pi^{2}}\right]}{r^{2}}(1 \\
&\left.\quad-\frac{4 \int d^{3} x\left[\frac{n_{q}\left(m_{u}+m_{d}+m_{s}\right)}{3}-\frac{4 a_{s}}{3 r}+b r+c-\frac{6 \Delta_{C F L}^{2} \mu^{2}}{n_{q} \pi^{2}}\right]}{r}\right)^{-\frac{1}{2}}
\end{aligned}
\end{equation}

Because of $|\overrightarrow{  {g}}|=\frac{  {GM}}{  {r}^{2}}$, the equivalent $  {G}$ under Natural Unit System becomes $  {G}=\frac{  {r}^{2}}{  {M}}|\overrightarrow{  {g}}| .$ Similarly, in non-CFL color-superconductive quark matter, the terms $-\frac{6 \Delta_{\mathrm{CFL}}^{2} \mu^{2}}{\mathrm{n}_{\mathrm{q}} \pi^{2}}$ in equation \eqref{e14} change into $-\frac{3 \Delta_{\mathrm{ds}}^{2} \mu^{2} \eta}{\mathrm{n}_{\mathrm{q}} \pi^{2}}-\frac{3 \Delta_{\mathrm{us}}^{2} \mu^{2} \eta}{\mathrm{n}_{\mathrm{q}} \pi^{2}}-\frac{3(1-2 \eta) \Delta_{\mathrm{ud}}^{2} \mu^{2}}{\mathrm{n}_{\mathrm{q}} \pi^{2}}$

\begin{figure}[H] %H为当前位置，!htb为忽略美学标准，htbp为浮动图形
\begin{center}%图片居中
\includegraphics[width=0.7\textwidth]{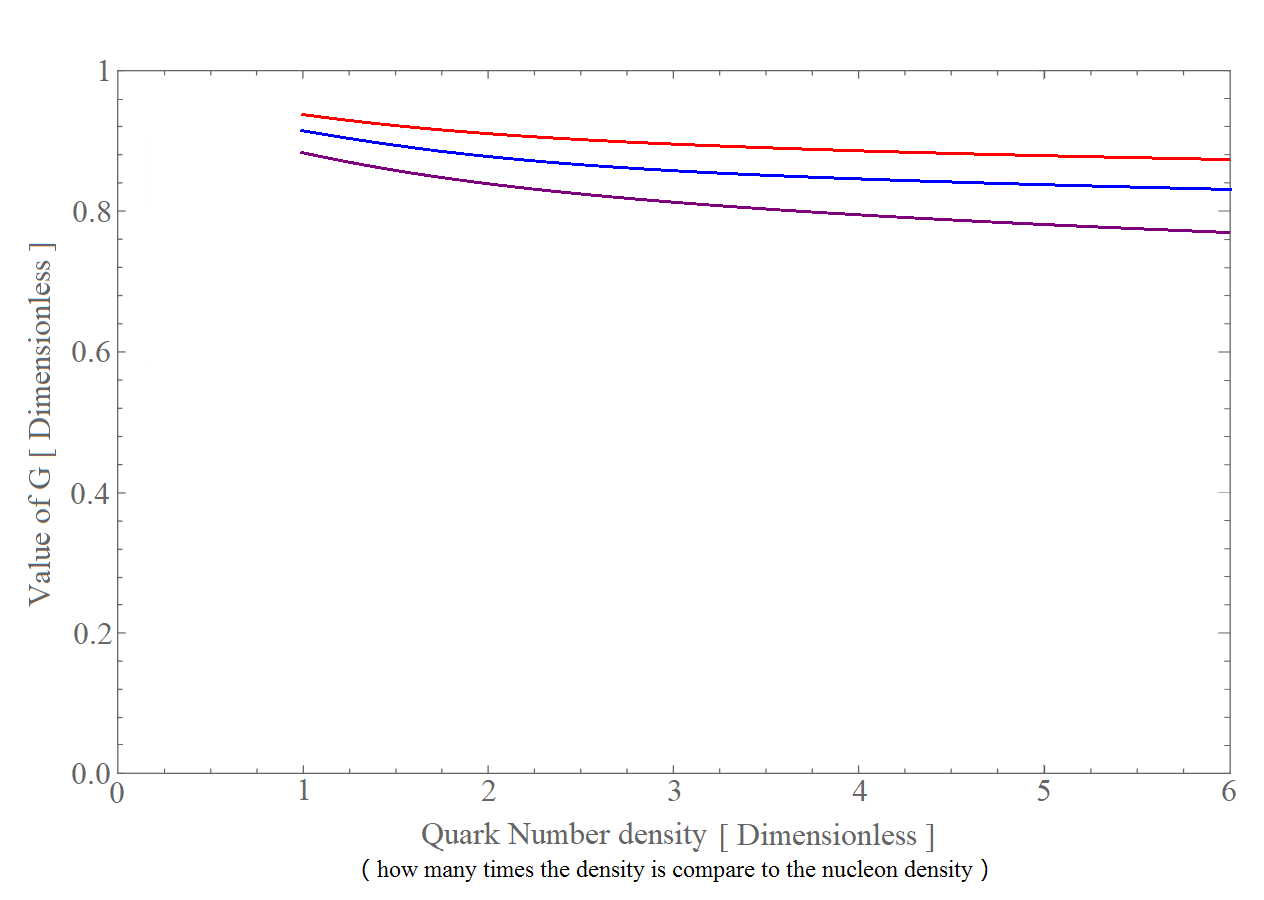} %插入图片，[]中设置图片大小，{}中是图片文件名
\end{center}
\caption{The equivalent G to be a function of quark number density, considering three cases when the mass of compact star is $2 \mathrm{M} \odot, 1 {M} \odot$, and $0.5 {M} \odot$ of lines from the bottom to the top}
 %最终文档中希望显示的图片标题
\label{Figure 1} %用于文内引用的标签
\end{figure}

\section{Discussion of Neutron star} \label{sec3}

A neutron star may contain a core of quark matter in general context, but it refers to a new forming neutron star which does not produce any quark matter in its core. Additionally, we will particularly concern the star’s inner part under high pressure, because it is the inner part where the quark nova explosion process will involve.
We denote the gravitational potential energy by $U$ and the kinetic energy of the spinning compact star by $K$. For simplicity, we can write the total energy (including internal energy) of neutron star ($NS$) and quark matter ($QM$) $E$ as:

\[
\begin{aligned}
{E}_{NS} &=\Omega_{{NS}}+\mu_{1} {N}_{1}+{S}_{1} {T}+{K}_{1}+{U}_{1} \\
{E}_{{QM}} &=\Omega_{{QM}}+\mu_{2} {N}_{2}+{S}_{2} {T}+{K}_{2}+{U}_{2}
\end{aligned}
\]

Subscripts $1$ and $2$ refer to the condition before and after the quark nova explosion, respectively. The generating of quarks  almost produces quarks and anti-quarks in pair, so for a random amount of matter that is transforming into quark matter , the difference between the numbers of quarks and anti-quarks can be considered as a conserved quantity before and after the quark nova. The new forming quark matter can be considered as entirely remaining  in the quark star or the quark matter core in neutron star, because the region matter crushes into quark matter should be under strong pressure, which means  it is inside the condensed star rather than on the surface and  become a jet in the quark nova explosion. Those analyses  above mean, for  an amount of matter, $\int \mathrm{d} x^{3} \mu_{1} N_{1}=\int \mathrm{d} x^{3} \mu_{2} N_{2}$. Thus, before and after the quark matter form, we have $\Omega_{{NS}}+\mu_{1} N_{1}+{S}_{1} T+{K}_{1}+{U}_{1}=\Omega_{{QM}}+\mu_{2} {N}_{2}+{S}_{2} {T}+{K}_{2}+{U}_{2}$, thus

\begin{equation}
\Omega_{{NS}}-\Omega_{{QM}}+\left({K}_{1}+{U}_{1}\right)-\left({K}_{2}+{U}_{2}\right)={T} \Delta {S}={Q}_{{Q} N}  \label{e15}
\end{equation}

Where $T$ here stands for the temperature of the location where  the process is happening. The releasing heat $Q_{QN}$ in this moment can be divided into two parts: one finally transformed into the kinetic energy of the outer layer matter of the star, and another finally transformed into the instant radiation. It is important to claim, because of the corrections from gravity potential and interaction terms, the equation \eqref{e14} does not include the whole contribution by quark nova  to the increase of the kinetic energy of the star outer layer and of the instant radiation. Here we use the simplification that the pressurein neutron star is mainly produced by baryons colliding and effects of relativistic degenerate matter. The pressure provided by momentum transfer in colliding or interacting is $P = \int \bar{k}n(p) v(p) \mathrm{d} p$, where $\bar{k}$ refers to the average momentum transfer in each colliding, $n(p)$ refers to the number density of particles with a momentum $p$ and $v(p)$ is their velocity. Therefore, we can write the total thermodynamic potential as

\begin{equation} \label{e16}
\begin{gathered}
\Omega_{  {NS}}=\frac{-4 \pi   {R}_{  {NS}}^{3}}{3}\left\{\frac {    { c } } { 8 \pi ^ { 2 } } \left\{\left(3 \pi^{2}   {n}_{  {q}}\right)^{\frac{1}{3}}\left\{\frac{2}{3}\left[\left(3 \pi^{2}   {n}_{  {q}}\right)^{\frac{1}{3}}\right]^{2}-  {m}_{  {q}}^{2}\right\} \sqrt{\left[\left(3 \pi^{2}   {n}_{  {q}}\right)^{\frac{1}{3}}\right]^{2}+  {m}_{  {q}}^{2}}\right.\right. \\
\left.+  {m}_{  {q}}^{4} \operatorname{arcsinh} \frac{\left(3 \pi^{2}   {n}_{  {q}}\right)^{\frac{1}{3}}}{  {~m}_{  {q}}}+\int \overline{  {k}} n(  {p})   {v}(  {p})   {d}   {p}\right\}(16)
\end{gathered}
\end{equation}
To estimate the number of collisions, we write the cross section

\begin{equation} \label{e17}
\frac{\mathrm{d} \sigma}{\mathrm{d} \Omega}=\frac{1}{64 \pi^{2} {E}_{{CM}}^{2}}|\mathcal{M}|^{2}, \quad {i} \mathcal{M}=\frac{-{ig}^{2}}{\overline{{k}}-{m}^{2}+{i} \varepsilon} 
\end{equation}

The new forming neutron stars usually spin rapidly, which provides a considerable contribution to the total energy

\begin{equation} \label{e18}
{K}_{1}=\frac{8 \pi \omega_{1}^{2}}{3} \int_{0}^{{R}_{{NS}}} {r}^{4} \rho_{1}^{2}({r}) \mathrm{d}r 
\end{equation}

The gravitational potential can be written as

\begin{equation} \label{e19}
{U}_{{g}}=-\int_{0}^{{R}_{{NS}}} {G} \frac{4 \pi {r}^{3} \rho_{1}}{3} 4 \pi r^{2} \rho_{1} \mathrm{d} r 
\end{equation}

Next, we consider the contribution of symmetry  energy. After the phase transformation process in quark nova, the remnant could be a quark star or a neutron star with a quark core if the quark star happened only a partial collapse that only crush the inner region of the star into quark matter. But in both cases, the transforming part of the star must be under great pressure, enabling it to crush into pure neutron state in the formation process of neutron star itself. In this case, comparing with symmetry energy in strange quark matter as equation \eqref{e7}, the symmetry energy in neutron star  \cite{dutra2014relativistic}\cite{hu2020effects} is written as

\begin{equation}\label{e20}
{E}_{{sym}({NS})}=\frac{{k}_{{F}}^{2}}{6 \sqrt{{M}_{{N}}^{* 2}+{k}_{{F}}^{2}}}+\frac{{g}_{\rho}^{2} \rho_{{B}}}{8\left({m}_{\rho}^{2}+2 \Lambda_{{V}} {g}_{\omega}^{2}{g}_{\rho}^{2} \omega^{2}\right)} 
\end{equation}
Here $k_F$ is the Fermi momentum of symmetric nuclear matter, $\Lambda=\frac{2}{3} {k}_{2}\left(\frac{{GM}}{{R}}\right)^{-5}$, and $R$ and $M$ are the neutron star’s radius and mass [5]. And comparing to that in quark matter, part of the symmetry energy would be released in quark nova explosion. 

\section{Transforming to Quark Matter} \label{sec4}

For simplicity, the process of quark nova explosion can be considered as three stages that producing energy for the explosion. Stage 1 is merely the collapse of the condensed star. This produces big pressure, leading to a rapid phase transformation into CFL phase in stage 2. And in stage 3, the producing energy releases as the kinetic energy of photons of strangelets and gives a contribution to the quark nova. In these stages the releasing heat, or the total energy contribution to the explosion, is written as ${Q}_{{QN}}={E}_{1}+{E}_{2}$. 

\subsection{Stage 1}
In this stage, $E_1$ comes from two parts: the change in gravitational potential energy and the change in the spinning kinetic energy to keep the conservation of angular momentum. We can derive $ E_1$ as:

\begin{equation} \label{e21}
\begin{gathered}
{E}_{1}=-\frac{8 \pi \omega_{1}^{2}}{3}\left[\frac{\left[\int_{0}^{{R}_{{NS}}} {r}^{4} \rho_{1}^{2} \mathrm{d} r\right]^{2}}{\int_{0}^{{R}_{{QM}}} {r}^{4} \rho_{2}^{2} \mathrm{d} r}-\int_{0}^{{R}_{{NS}}} {r}^{4} \rho_{1}^{2} {\mathrm{d}r}\right]+\int_{0}^{{R}_{{QM}}} {G} \frac{4 \pi {r}^{3} \rho_{2}}{3} 4 \pi {r}^{2} \rho_{2} \mathrm{d} r\\
-\int_{0}^{{R}_{{NS}}} {G} \frac{4 \pi {r}^{3} \rho_{1}}{3} 4 \pi r^{2} \rho_{1} \mathrm{d} r   
\end{gathered}
\end{equation}
where $\omega_1$ here is angular velocity of the spinning neutron star before quark nova explosion and the angular velocity after that is determined by conservation of angular momentum; $\rho_2$ is the density in quark matter. Because our current focus is the moment at which quark nova explosion happens, and when the forming quark matter has not reached a balance state of density distribution yet, we can simplify the case by assuming the density has not changed after its formation. In other words, $\rho_2$ is still on the boundary condition. As previously described  \cite{lugones2002color}, the condition of the phase transformation to CFL to happen determines that $\frac{\rho_{2}}{\rho_{1}}=\frac{{n}_{2}}{{n}_{1}}=\frac{7}{3}, {n}_{1}={n}_{{B}}, \mathrm{n}_{2}=\frac{1}{3} {n}_{{q}}$. During this collapsing stage, the particles are sped up by gravity and therefore heat the forming quark matter and the outer part of the neutron star. During the compact star which remains after quark nova  cooling down to the temperature before quark star’s explosion, this part of energy is turning to the form of radiation. Also, some of the particles will bounce back and speed up the escaping of the outer layer of the star.

\subsection{Stage 2} 

In the CFL phase, the thermodynamic potential of whole CFL quark matter ($m_n=939$ MeV) \cite{lugones2002color}.

\begin{equation}\label{e22}
\Omega_{{CFL}}=-4 \pi \int_{0}^{{R}_{{QM}}} {r}^{2}\left(\frac{3 \mu({r})^{4}}{4 \pi^{2}}-\frac{3{m}_{{s}}^{2} \mu({r})^{2}}{4 \pi^{2}}+\frac{3 \Delta_{{CFL}}^{2} \mu({r})^{2}}{\pi^{2}}+\frac{{m}_{{s}}^{2}{m}_{{n}}^{2}}{12 \pi^{2}}-\frac{\Delta_{{CFL}}^{2} {m}_{{n}}^{2}}{3 \pi^{2}}-\frac{{m}_{{n}}^{4}}{108 \pi^{2}}\right) \mathrm{d} r 
\end{equation}
For random amount of quark matter, $\mu(r)$ satisfies $\frac{\partial \Omega}{\partial \mu}=0$. Here we use
\[
\mu({r})^{2}=\frac{{m}_{3}^{2}}{6}-\frac{2 \Delta_{{CFL}}^{2}}{3}+\left(\left(\frac{{m}_{8}^{2}}{6}-\frac{2 \Delta_{{CFL}}^{2}}{3}\right)^{2}+\frac{4 \pi^{2}}{9}\left(\frac{\rho_{{c}}\left(3+{ar}^{2}\right)}{3\left(3+{ar}^{2}\right)^{2}}-{B}\right)\right)^{\frac{1}{2}}
\]
and $\rho_c$ refers to the central density while $a$ is a function of $\rho_c$  \cite{rocha2019exact}. 

In the forming of Color Flavor Locked phase,  the free energy gained from CFL pairing is greater than that cost by maintaining equal quark number densities  \cite{alford2001minimal}. Thus, the unpaired quark matter would spontaneously transform into CFL phase. However, if the effective strange quark mass $m_s$ is large while the gap $\triangle$ \cite{alford2004gapless} is small here, the mass of strange quark might cost more energy than the symmetry energy  \cite{alford2001minimal}. The critical point depends on whether the symmetry energy of quark matter  \cite{cai2012nuclear} to satisfy that ${E}_{{sym}({QM})} \geq {m}_{{s}}$ we regard

\begin{equation}\label{e23}
\frac{1}{2} \frac{v^{2}}{\sqrt{v^{2}+m^{2}}} \frac{3 n_{B}-n_{s}}{3 n_{B}}\left(\frac{n_{d}-n_{u}}{n_{d}+n_{u}}\right)^{2} \geq m_{s} 
\end{equation}
as the boundary condition.

As for the rest part of $E_2$, the only difference is the quark interacting term due to the conversation of electric charge. Using the effective “(anti) quark” model in part \ref{sec2}, with the parameters in Cornell potential in different conditions is shown in  \cite{pathak2020parameterisation}. The average distance $r$ in CFL quark matter $ {r}={r}_{0}\left(\frac{{n}_{1}}{{n}_{2}}\right)^{\frac{1}{3}}$ where $r_0$ is the scale of neutron. To compare with CFL state , the quark interaction potential is included in nuclei mass in nuclear matter in neutron star . The difference of this potential energy in the two cases  per baryon number is roughly 0.3GeV due to the different average distance. So according to above all, consider the symmetry energy mentioned in Part \ref{sec3} , using $\Omega_NS$ and $E_sym$ given by Part \ref{sec3}  we have

\begin{equation}\label{e24}
{E}_{2}=\Omega_{{NS}}+{n}_{{B}}\left[{m}_{{n}}-\left({m}_{{u}}+{m}_{{d}}+{m}_{{s}}\right)\right]+{E}_{{sym}({NS})}-\Omega_{{QM}}-\frac{{n}_{{q}}}{2} {V}  
\end{equation}
Here $E_{\operatorname{sym}(Q M)}$ is the $E_{\text {sym }}\left(n_{B}, \delta, n_{s}\right)$ in equation \eqref{e7}.

\subsection{Stage 3}

Considering the effect of the gravity of the compact star, the releasing energy before and after the two stages in the form of kinetic energy of outer layer and instant radiation can be written as

\begin{equation}\label{e25}
{K}_{\text {outerlayer }}+\varepsilon_{\text {radiation }}=-\frac{{GM}_{{NS}}}{{R}_{{NS}}}\left({m}_{\text {outerlayer }}+\varepsilon_{\text {radiation }}\right)+{Q}_{{QN}} 
\end{equation}

Actually, the outer layer of the star basically stays in the nebula relatively far from the compact star or is blown away relatively slowly; therefore, in general the term $K_\text{outerlayer}$ can be ignored in calculation and comparison with real observation data. As a result, it is simplified that

\begin{equation}\label{e26}
\begin{gathered}
\varepsilon_{\text {radiation }}=-\frac{{GM}_{{NS}}}{{R}_{{NS}}}\left({m}_{\text {outerlayer }}+\varepsilon_{\text {radiation }}\right)-\frac{8 \pi \omega_{1}^{2}}{3}\left[\frac{\left[\int_{0}^{{R}_{{NS}}} {r}^{4} \rho_{1}^{2} {dr}\right]^{2}}{\int_{0}^{{R}_{{Q} M}} {r}^{4} \rho_{2}^{2} \mathrm{d} r}-\int_{0}^{{R}_{{NS}}} {r}^{4} \rho_{1}^{2} {dr}\right] \\
+\frac{8}{9} \pi^{2} \rho_{2}^{2} {GR}_{{QM}}^{6}-\frac{8}{9} \pi^{2} \rho_{1}^{2} {GR}_{{NS}}^{6}+\Omega_{{NS}}+{n}_{{B}}\left[\mathrm{m}_{{n}}-\sum_{{q}={u}, {d}, {s}} {m}_{{q}}\right]+{E}_{{sym}({NS})} \\
- E_{\operatorname{sym}(Q M)}-\Omega_{{QM}}-\frac{{n}_{{q}}}{2} {V} 
\end{gathered}
\end{equation}

Here $\varepsilon$ refers to the difference of Cornell potential per neutron between neutron star matter and quark matter which is given by $\varepsilon=m_{n}-\sum_{q=u, d, s} m_{q}-\frac{3}{2} V$. The strength of radiation is mainly related to  the mass of the part of neutron star which is involved in the quark nova. When only a part of the initial neutron star involves in the quark nova, the energy of radiation can be partially absorbed by the outer crust of nuclei matter in the remaining compact star. In this hybrid star case \cite{blaschke1999dynamical} \cite{buballa2004quark} \cite{shao2011evolution} using equation \eqref{e2} about thermodynamic potential in Part \ref{sec3} the energy releasing quantity is given in Figure 2.

Comparing with the observed supernova events, the results in this table can  support the view that quark nova explosions can be as common as neutron stars with core of quark matter \cite{annala2019quark}, and  the energy released in this process could possibly fix the missing energy problem in need in the ordinary supernova model. 

\begin{figure}[H] %H为当前位置，!htb为忽略美学标准，htbp为浮动图形
\begin{center} %图片居中
\includegraphics[width=0.7\textwidth]{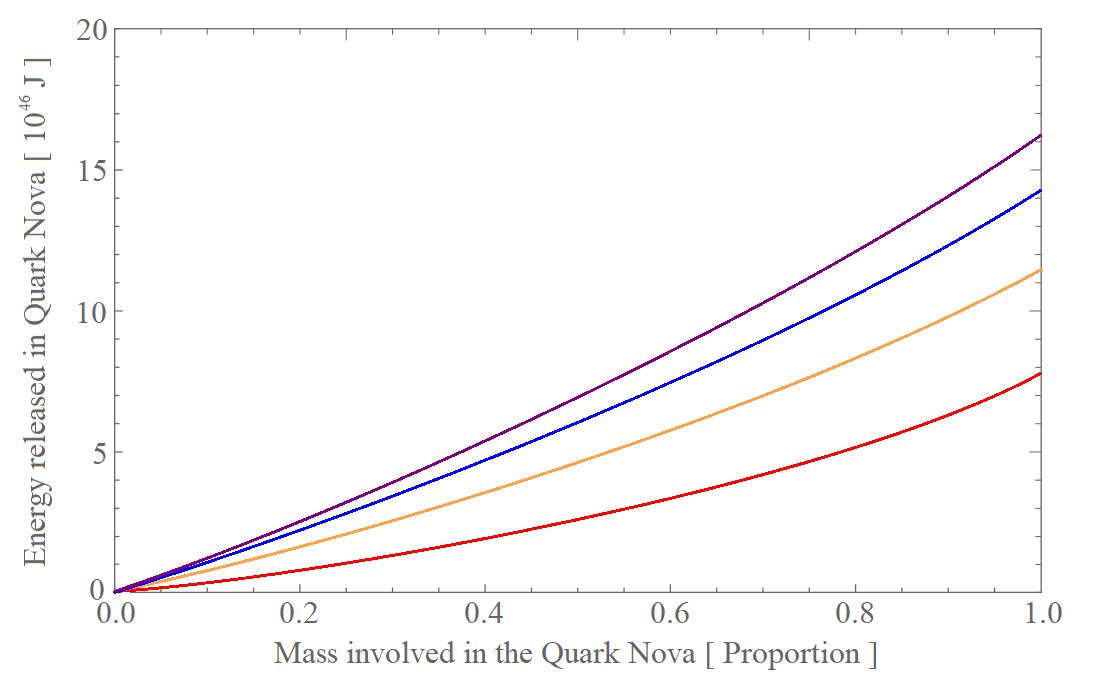} %插入图片，[]中设置图片大小，{}中是图片文件名
\end{center}
Figure 2: Energy releasing in quark nova explosion with different initial neutron star mass and different proportions of matter that involves in this process. The initial neutron star massof lines from the bottom to the top are $1.5M\odot$, $2M\odot$, $2.5M\odot$, and $3M\odot$(if possible \cite{burgio2002maximum}). One thing to clarify is the initial mass mentioned above does not mean the final mass of the remaining compact star, because of the reduction in the quark nova explosion. Therefore, the “initial mass” is not required to be in the stable limit of neutron stars for it could collapse in the quark star process immediately. After the collapsing, the quark star should satisfy the Chandrasekhar limit \cite{halder2021chandrasekhar}. Former researches (\cite{li2011nearby} \cite{richardson2002comparative}) have analysed the data of supernova, And it can be summarized in the following table:

%最终文档中希望显示的图片标题
\label{Figure 2} %用于文内引用的标签
\end{figure}

\begin{tabular}{|p{1cm}<{\centering}|p{2.5cm}<{\centering}|p{2cm}<{\centering}|p{2cm}<{\centering}|p{2cm}<{\centering}|} 
\hline Type &  Average peak  absolute magnitudeb & Approximate energy (foe)c & Days to peak luminosity & Days from peak to 10\% luminosity \\
\hline Ia & $-19$ & 1 & approx. 19 & around 60 \\
\hline Ib/c (faint) & around-15 & $0.1$ & $15-25$ & unknown \\
\hline Ib & around $-17$ & 1 & $15-25$ & $40-100$ \\
\hline Ic & around $-16$ & 1 & $15-25$ & $40-100$ \\
\hline Ic (bright) & to $-22$ & above 5 & roughly 25 & roughly 100 \\
\hline II-b & around $-17$ & 1 & around 20 & around 100 \\
\hline II-L & around -17 & 1 & around 13 & around 150 \\
\hline II-P (faint) & around -14 & $0.1$ & roughly 15 & Unknown \\
\hline II-P & around -16 & 1 & around 15 & Plateau then around 50 \\
\hline IInd & around -17 & 1 & $12-30$ or more & 50-150 \\
\hline In (bright) & to $-22$ & above 5 & above 50 & above 100 \\
\hline
\end{tabular}

Although most part of energy released from quark nova is in the form of neutrino, and in the large initial neutron star mass cases, the proportion of mass involved in the quark nova might not be very high due to the explosion would blow away part of the mass, it could still provide enough energy to release radiation \cite{ouyed2005ultra} and speed the outer layer of the supernova remnant to escape from gravity. One thing should also be mentioned is that if the initial neutron star matter is too small or the supernova is not powerful enough to give the first crush for the neutron star matter to reach the critical condition(like the case in type Ia supernova ), the quark nova process will  not happen. The critical condition in compact star is discussed above and more generally discussion for non compact star case would be given in the following part.

\section{Strangelets Escaped from Compact Stars} \label{sec5}

In the end of the last section, we discussed the QCD phase transformation in compact star. In this section we will generalize our fields to non-compact star case. A well-known problem of strange quark matter theory is the existence and effect of strangelets has not been observed yet. The reason could be the overestimate of the quantity of strangelets, the exclusion ofreaction producing strange quark matter being a chain reaction, or the failure of reaction in low energy state or low chemical potential environment outside the compact stars. Our present study mainly focuses on the latter two possibilities.

Since the total energy E is conserved during the process, $\Omega_{ {n}}+\mu_{ {n}}  {N}_{ {n}}+ {S}_{ {n}}  {T}+ {U}_{ {n}}=\Omega_{ {q}}+\mu_{ {q}}  {N}_{ {q}}+ {S}_{ {q}}  {T}+ {U}_{ {q}}$, and if the phase transform reaction to CFL state  could possibly be an exergonic reaction beyond the environment in compact star, it should satisfy that $\int \mu_{{n}} {N}_{{n}} \mathrm{d}x^{3}=\int \mu_{{q}} {N}_{{q}} \mathrm{d} {x}^{3}$ implies $ {Q}= {U}_{ {n}}+\int- {p}_{ {n}}  {dV}_{ {n}}- {U}_{ {q}}-\int- {p}_{ {q}}  {d}  {V}_{ {q}}>0$. Additionally, if $Q$ is larger than the energy in need to crush the nearby ordinary matter together, reaching the condition for this reaction to happen  ,the reaction  could possibly be a chain reaction. Here we will make an estimate under the assumption of a simplest situation for nuclei and quark matter in a sphere isotropic from its center. In this case, first we derive the pressure by perturbation approach and then give the correction B by MIT bag model  \cite{harko2015wormhole}. Denote the baryon number involved in the reaction as $n_B$. As a small amount of matter, the macroscopic effects such as gravitational potential can be ignored but the  microscopic effects of particles are now under concerned. For example, because the quark matter is electrically neutral, a random unit charge of a distance $r$ from the center has an electric potential energy of $-\frac{e^{2}}{2 r}$, and the rest of particles can be equivalently regarded as another unit charge with the opposite charge of the same distance $r$ to the center at the opposite position in the sphere. Integrate with respect to $r$ and then we obtain the electromagnetic binding energy $-\frac{3 n_{B} e^{2}}{16 \pi r_{0}}$ ($r_0$ is radius of the sphere). Ignoring gravity we regard the particles as uniformly distributed. Denoting $\bar{k}$ as the average momentum transfer, we can derive the total energy as 

\begin{equation}\label{e27}
 {E}_{\text {nuclear }}=- {n}_{ {B}} \overline{ {v}}_{ {n}} \overline{ {k}}_{ {N}}+ {n}_{ {B}}\left[ {m}_{ {n}}-\sum_{ {q}= {u},  {d},  {s}}  {m}_{ {q}}\right]-\frac{3  {n}_{ {B}}  {e}^{2}}{16 \pi  {r}_{ {N}}}+\int \mu  {Ndx}^{3} 
\end{equation}

\begin{equation}\label{e28}
 {E}_{ {CFL}}=-\left( {n}_{ {q}} \overline{ {v}}_{ {q}} \overline{ {k}}_{ {ud}}+ {n}_{ {q}} \overline{ {v}}_{ {q}} \overline{ {k}}_{ {ds}}+ {n}_{ {q}} \overline{ {v}}_{ {q}} \overline{ {k}}_{ {us}}\right)+ \frac{n_q}{2}\left(-\frac{4  {a}_{ {s}}}{3  {r}}+ {br}+ {c}\right)-\frac{3  {n}_{ {B}}  {e}^{2}}{16 \pi  {e}_{ {Q}}}+\int \mu  {Ndx}^{3}+ {BV} 
\end{equation}

$V$ is the volume of strangelet. Therefore, the criterion for being an exergonic reaction can be written as

\begin{equation}\label{e29}
\begin{gathered}
 {Q}_{1}=- {n}_{ {B}} \overline{ {n}}_{ {n}}  {Q}_{ {N}}+ {n}_{ {B}}\left[ {m}_{ {n}}-\sum_{ {q}= {u},  {d},  {s}}  {m}_{ {q}}\right]-\frac{3  {n}_{ {B}}  {e}^{2}}{16 \pi  {r}_{ {N}}}+\left( {n}_{ {q}} \overline{ {v}}_{ {q}} \overline{ {k}}_{ {ud}}+ {n}_{ {q}} \overline{ {v}}_{ {q}} \overline{ {k}}_{ {ds}}+ {n}_{ {q}} \overline{ {v}}_{ {q}} \overline{ {k}}_{ {us}}\right) \\
-\frac{n_q}{2} \left(-\frac{4  {a}_{ {s}}}{3  {r}}+ {br}+ {c}\right)+\frac{3  {n}_{ {B}}  {e}^{2}}{16 \pi  {r}_{ {Q}}}- {BV}>0 
\end{gathered}
\end{equation}

Next, we will discuss how much energy the process costs to crush ordinary matter into the critical condition for CFL phase transformation.

First, we consider the case when energy is inputted from outside and focus on a small region. In this case, the energy in need would include two main parts, that to speed up neutrons and that to crush it into the high-pressure condition. When two beams of extreme relativistic particles with a total kinetic energy $K_0$ collide, if in each beam the particles behind are faster than the particles in front, the particles will be focused and strongly crushed at the moment they collide. This will create an instant extreme high-pressure environment and reach the phase transforming critical condition with energy $E_\text{nuclear}$, and energy for this process to happen is entirely provided by initial kinetic energy $K_0$. 

Then, turn to the case when CFL phase transformation has happened nearby. In this case, the energy released by the phase transformation will provide the initial kinetic energy for continuing  of this reaction. The energy density on the surface of the strangelet is $\sigma=\frac{{Q}_{1}}{4 \pi {R}^{2}}$ with $R$ being the radius of strangelet. And the critical condition requires at least $\frac{ {Q}_{1}  {r}^{2}}{4 \pi  {R}^{2}} \geq 2  {~m}_{ {s}}+ {V}$, where $r$ is the radius of nuclei.

The criterion for the process being sustained naturally can be written as

\begin{equation}\label{e30}
 {Q}_{2}=\frac{ {Q}_{1}  {r}^{2}}{4 \pi  {R}^{2}}-2  {~m}_{ {s}}- {V}>0 
\end{equation}

But to be a chain reaction, it requires that the thickness of the interacting layer should not become thinner and thinner in the process of passing. Thus the environmental energy density must be high enough for the quark matter to form, and the requirement is

\begin{equation} \label{e31}
-\left( {n}_{ {q}} \overline{ {v}}_{ {q}} \overline{ {k}}_{ {ud}}+ {n}_{ {q}} \overline{ {v}}_{ {q}} \overline{ {k}}_{ {ds}}+ {n}_{ {q}} \overline{ {v}}_{ {q}} \overline{ {k}}_{ {us}}\right)+ \frac{n_q}{2} \left(-\frac{4  {a}_{ {s}}}{3  {r}}+ {br}+ {c}\right)-\frac{3  {n}_{ {B}}  {e}^{2}}{16 \pi  {r}_{ {Q}}}+ {BV}<0
\end{equation}
The region for it to be chain reaction is shown in Figure 3.

\begin{figure}[H] %H为当前位置，!htb为忽略美学标准，htbp为浮动图形
\begin{center}%图片居中
\includegraphics[width=0.7\textwidth]{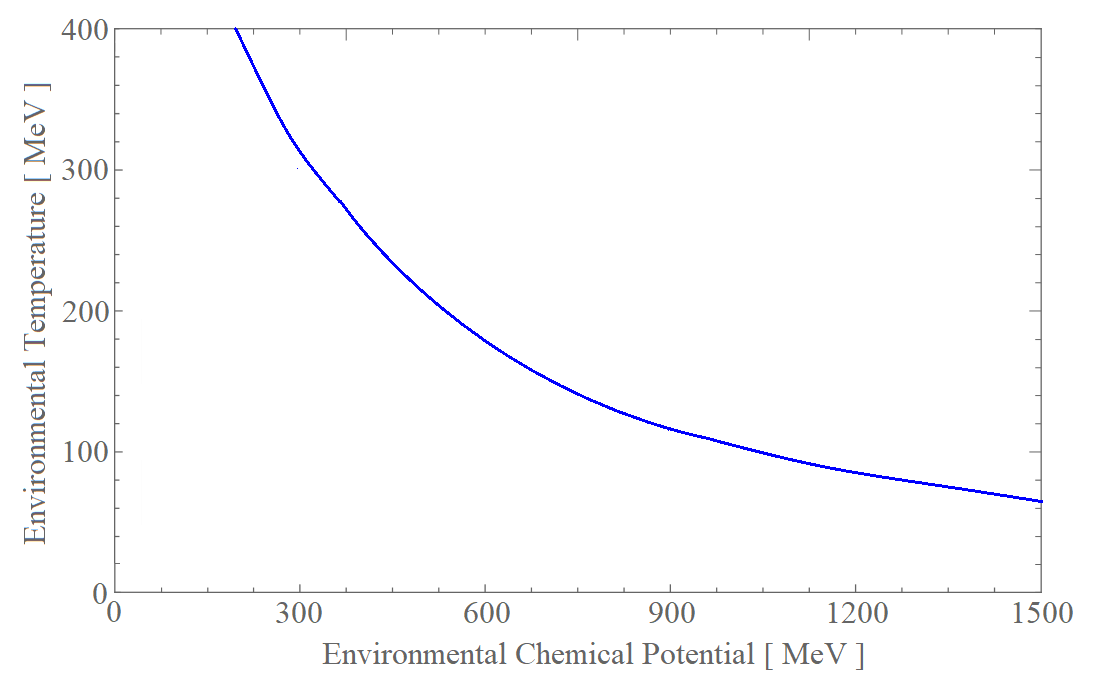} %插入图片，[]中设置图片大小，{}中是图片文件名
\end{center}
Figure 3: The environmental temperature in need with different environmental chemical potential. The result shows this reaction might not be a chain reaction in common condition outside compact stars or particle colliders.  %最终文档中希望显示的图片标题
\label{Figure 3} %用于文内引用的标签
\end{figure}

\section{Conclusion} \label{sec6}

In this paper, we use the view of QCD thermodynamics to analyse the quark nova process and give a equation of the amount of releasing energy dependent on the mass distribution of the forming neutron star. With several simplifications such as the assumption of the density in the core or inner part of neutron star as roughly a constant, the pressure in neutron star as being provided entirely from particles colliding and scattering, and all the forming quark matter as in CFL phase, we give the result in figure 2. More generally, the strangelets outside compact stars are also analysed in Part \ref{sec4}, shows the requirements to produce them . The results in Part \ref{sec5} suggests the reaction to produce strangelets  should not be a chain reaction under the common condition outside compact stars.

\normalem    % 防止参考文献出现下划线
%\footnotesize % 调了顺序后，参考文献字体变大，回归正常字体

%\bibliographystyle{IEEEtran}
%\bibliography{IEEEabrv,mybibfile}

\bibliography{PDE1}{}

\bibliographystyle{unsrt}

\end{document}